\documentclass[plb]{elsarticle}

\usepackage{lineno,hyperref}
\modulolinenumbers[5]
\usepackage[T1]{fontenc} 
\usepackage{amsmath,amssymb,graphicx,bbm,mathrsfs,nicefrac}
\usepackage{graphicx}
\usepackage{tikz}

\def\bsp#1\esp{\begin{split}#1\end{split}}

\newcommand{\be}{\begin{equation}} 
\newcommand{\ee}{\end{equation}}  
\newcommand{\bea}{\begin{eqnarray}}  
\newcommand{\eea}{\end{eqnarray}}  
\def\bpm{\begin{pmatrix}}
\def\epm{\end{pmatrix}}

\def\bsp#1\esp{\begin{split}#1\end{split}}
\def\spa#1.#2{\left\langle #1 \, #2 \right\rangle}
\def\spb#1.#2{\left[ #1 \, #2 \right]}
\def\spab#1.#2.#3{\left\langle #1 |#2| #3 \right]}
\def\spaa#1.#2.#3.#4{\left\langle #1 |#2 |#3 | #4 \right\rangle}
\def\spbb#1.#2.#3.#4{\left[ #1 | #2 | #3 | #4 \right]}

\newcommand\lsim{\mathrel{\rlap{\lower4pt\hbox{\hskip1pt$\sim$}}
    \raise1pt\hbox{$<$}}}
\newcommand\gsim{\mathrel{\rlap{\lower4pt\hbox{\hskip1pt$\sim$}}
    \raise1pt\hbox{$>$}}}

\newcommand{\captionfonts}{\small}

\newcommand{\approptoinn}[2]{\mathrel{\vcenter{
  \offinterlineskip\halign{\hfil$##$\cr
    #1\propto\cr\noalign{\kern2pt}#1\sim\cr\noalign{\kern-2pt}}}}}

\makeatletter  
\long\def\@akecaption#1#2{%
  \vskip\abovecaptionskip
  \sbox\@tempboxa{{\captionfonts #1: #2}}%
  \ifdim \wd\@tempboxa >\hsize
    {\captionfonts #1: #2\par}
  \else
    \hbox to\hsize{\hfil\box\@tempboxa\hfil}%
  \fi
  \vskip\belowcaptionskip}
  
\makeatother   

\begin{document}

\begin{frontmatter}
\title{ On the compatibility of the diboson excess with a gg-initiated composite sector}

 \ead{v.sanz@sussex.ac.uk}
\author{Ver\'onica Sanz}
\address{Department of Physics and Astronomy, University of Sussex, Brighton BN1 9QH, UK}

\begin{abstract}
In this paper we propose that recent results by ATLAS and CMS searching for heavy resonances decaying into bosons could be a first hint of  a new sector of pure-gauge confining physics, possibly linked to the origin of the Higgs as a Composite Higgs. The lightest resonances (glueballs) of this new sector would be neutral, spin-zero and -two, and their behaviour would resemble that of a radion and a massive graviton of extra-dimensions. We outline how 13 TeV LHC data could be used to improve sensitivity on this scenario, as well as future characterization during the 13 TeV LHC run.
\end{abstract}

\end{frontmatter}

\section{Introduction}

Searches for heavy resonances decaying into a pair of bosons performed by CMS and ATLAS~\cite{CMSVV,ATLASVV,ATLASVl,CMSVl,ATLASV2l} shows tantalizing hints towards the existence of a new resonance at a mass of around 2 TeV, a possibility which has created quite some excitement~\cite{new,new2,new3,new4,new5,new6,new7,new8,new9,new10,new11,new12,new13,new14,new15,new16}. 

In this paper we provide an alternative interpretation in terms of a new strong sector, possibly linked to the origin of the Higgs particle as a composite state. We will consider new states, singlet under the SM interactions, which can be produced and decay through their coupling to the stress-energy tensor. An example of such a theory is a new pure gauge sector which undergoes confinement at energies around the TeV scale. The spectrum of this theory contains glueballs, with spin-zero and spin-two resonances at the bottom of the spectrum~\cite{glueballs-review,glueballs-review2,glueballs-review3,glueballs-review4,glueballs-review5}. Focusing on the low-lying, conceivably narrow states, we concentrate in the scenario where the new states couple to gluons through anomalies, and decay predominantly to massive vector bosons or Higgses.  
  

 \section{Glueballs: Theoretical aspects} \label{sec1}

Consider a new non-abelian gauge sector, e.g. a $SU(n)$ gauge group, which undergoes confinement leading to a low energy spectrum of glueballs. Glueballs are bound states of gauge fields and their behaviour have been studied both in the case of QCD and of more general gauge theories. The ordering of states can be understood by examining the interpolating operators of minimal canonical dimension~\cite{Kuti}, a prescription which lattice simulations seem to confirm~\cite{glueballs-review}. 

With this prescription, the lightest states would then correspond to those generated by the lowest dimensional singlet operator, namely a dimension-four operator $Tr F_{\mu\nu} F_{\alpha\beta} $, 
which generates glueballs with quantum numbers $J^{PC}=0^{++}$, $2^{++}$, $0^{-+}$, $2^{-+}$~\footnote{Note that these quantum numbers could also be achieved within {\it oddballs}~\cite{llanes}}.

The next level of resonances would be associated with the dimension-five operator, $Tr F_{\mu\nu} D_{\rho} F_{\alpha\beta}$ leading to resonances  with quantum numbers $J^{PC}=1^{++}$ and $3^{++}$. One could continue this procedure to classify resonances by examining dimension-six and higher operators generated by gauge fields. 

In the following we are going to focus on the lowest resonances, of spin-zero and spin-two. Lattice studies on $SU(n)$ gauge theories find that the lowest resonances correspond to $PC=++$, hence we will denote them by
\bea
\phi : J^{PC}= 0^{++} \textrm{ and } G_{\mu\nu} : J^{PC}= 2^{++} \ .
\eea
Determining the separation between the scalar ($0^{++}$) and tensor states ($2^{++}$) is a difficult task in lattice gauge theories. Lattice results in a pure glueball calculation indicate that the tensor mode is about 60\% heavier than the scalar, even in the large-$n$ limit.  Nevertheless, this result will likely change once the pure gauge theory is coupled to the Standard Model.  In the following, I will consider these two lightest  states as two distinct possibilities for the lightest state in a glueball spectrum.

\begin{figure}[h!]
\begin{center}
\includegraphics[width=.45\textwidth]{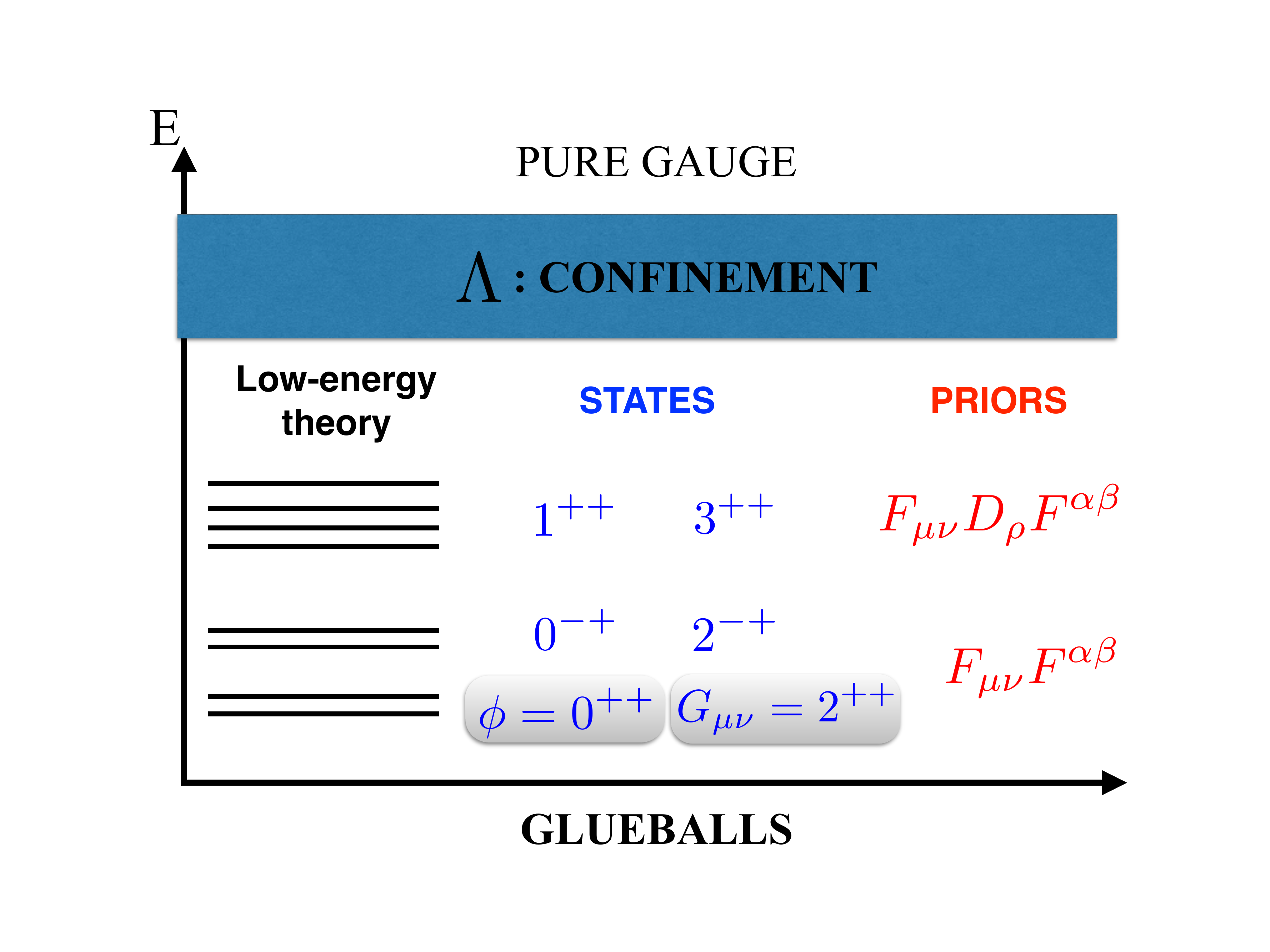}
\includegraphics[width=.45\textwidth]{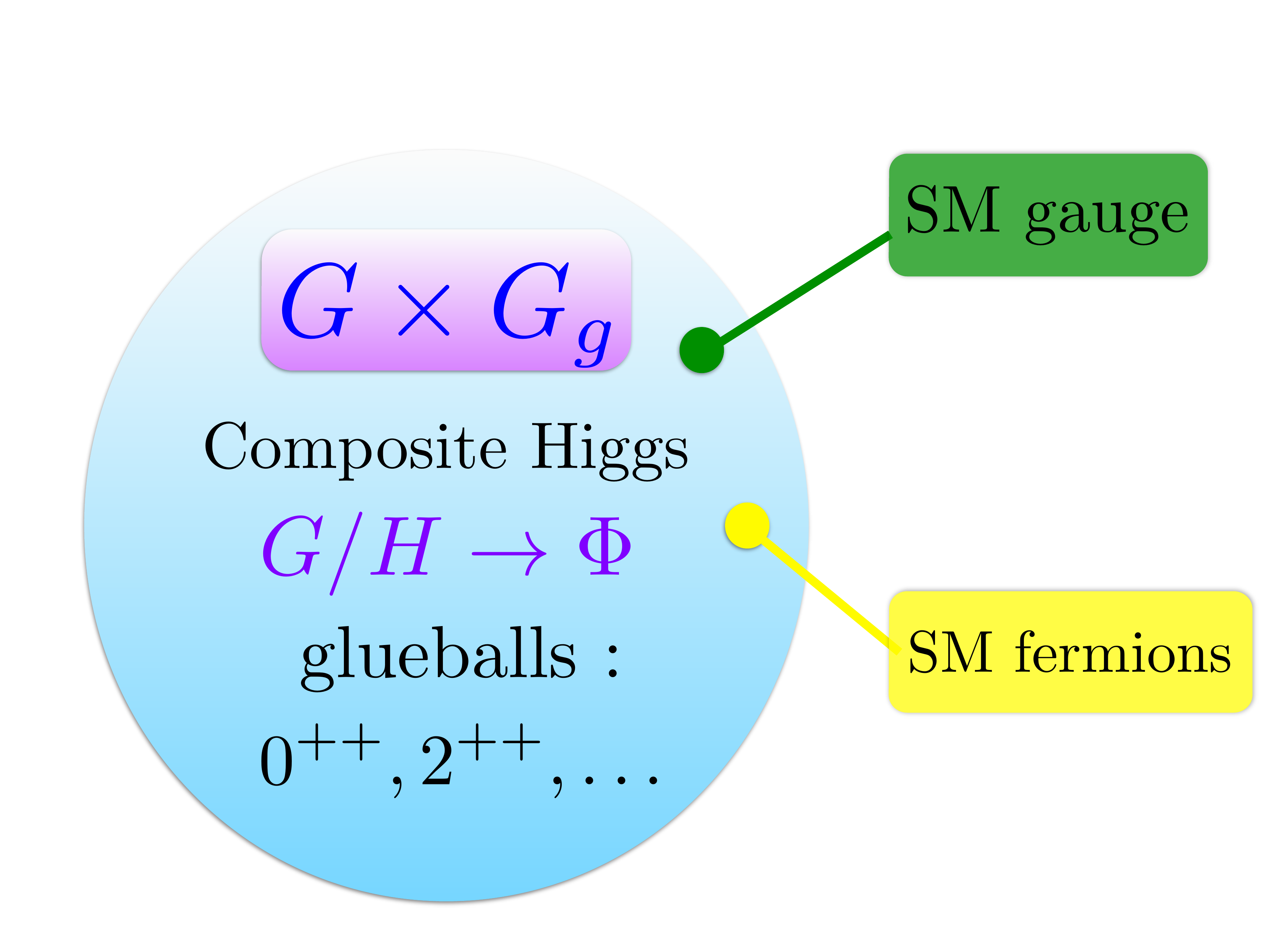}
\label{fig:sketch}
\caption{({\it Left: }) Spectrum of glueballs in a pure gauge theory, including the prior operators. The two lightest states are singled out as $\phi$ ($0^{++}$) and $G_{\mu\nu}$ ($2^{++}$). ({\it Right:}) The new sector of strong interactions exhibits a global symmetry $G$ broken spontaneously by $G_g$ strong dynamics.}
\end{center}
\end{figure}
 
The resonances $\phi$ and $G$ propagate as a Klein-Gordon and a Fierz-Pauli~\cite{fierz-pauli} fields. The  Fierz-Pauli Lagrangian describes a massive spin-two field, a rank-two symmetric and traceless tensor. Additionally, a positive-energy condition must be satisfied, see Ref.~\cite{gravitonus} and references therein. 
\subsection{Generic couplings of the spin-zero and two states} 

Contrary to the case of QCD, a pure gauge theory has no global (chiral) symmetries which would be broken by the confinement dynamics. On the other hand, space-time symmetries can be broken by confinement. For example, glueballs break spontaneously scale invariance of the gauge theory. Hence, the lightest spin-zero resonance could play the role of a dilaton, the Goldstone boson of the spontaneous breaking of scale invariance. In this case, the couplings of the dilaton resonance are of the form $\frac{\phi}{\Lambda} \partial_{\mu} J^{\mu}$, 
 where $J^{\mu}$ is the global current whose spontaneous breaking at the scale $\Lambda$ leads to the emergence of the Goldstone boson $\phi$. The global current is given by  $J^{\mu}= T^{\mu\nu} v_{\nu}$, 
 where $v_\mu$ is the generator of dilatation symmetry $v_{\nu} = \lambda x_{\nu}$, which then implies that the dilaton couples to trace of the stress tensor $T=T_\mu^\mu$,
 \bea
- \frac{a_i}{\Lambda} \phi \, T_i  \label{r1}
 \eea
 where $\Lambda$ is the symmetry breaking scale and the index $i$ refers to species, i.e. $i=$ $\Phi, \gamma, g, \ldots$ In our set-up the dimensionless parameter $a_i$ encodes the degree of compositeness of species $i$, with an order one value indicating a large mixture with the composite sector. Note that this coupling vanishes for massless gauge bosons (gluon and photon) as $Tr T_{\mu\nu}^{\gamma, g} =0$, but a coupling would be induced nevertheless at one-loop through the anomaly~\cite{real-radion,Witek-Jiji,diHiggs},
 \bea
- \frac{a_{g,\gamma}}{\Lambda} \phi F_{\mu\nu} F^{\mu\nu} \label{r2} \ ,
 \eea
 where $F$ here denotes the gluon or photon field-strength.

We encounter a similar situation for the spin-two state $G_{\mu\nu}$ whose properties are derived from diffeomorphism invariance, broken spontaneously by the gauge dynamics.
Indeed, a massless spin-two object $\theta^{\mu\nu}$ is conserved, $\partial_{\mu} \theta^{\mu\nu}=0$ in the absence of breaking. As in the $0^+$ case, $\theta_{\mu\nu}$ it couples to a conserved current $\partial_{\mu} J^{\mu}=0$. The breaking of this diffeomorphism invariance can be parametrized by $\partial_{\mu} \theta^{\mu\nu}=v^{\nu}$, where $v^{\mu}$ corresponds to a massive vector field, which is {\it eaten} by the massless spin-two field~\cite{massives2,massives22,massives3}. When joining together, the spin-two massless field and the massive vector will lead to the massive spin-two state $G_{
\mu\nu}$. As long as the composite sector preserves Lorentz, gauge and CP invariance, the coupling of the massive spin-two resonance to {\it two} SM particles will be given by~\cite{gravitonus}
 \bea
- \frac{b_i}{\Lambda} \, G_{\mu\nu} \, T^{\mu\nu}_i \ ,
 \eea
 where $T^{\mu\nu}_i$ corresponds to the stress tensor of species $i$. Studies of couplings to the tensor state to SM particles have been done in the context of glueballs in QCD~\cite{stephan}.  Note that these analyses differ from ours in which the tensor two-point function is treated as containing a massless state, and the constraints from diffeomorphism invariance are not included.

\subsection{A Set-up for Glueballs and a Composite Higgs}
 
 In a scenario where EWSB is due to strong dynamics, such as Composite Higgs scenarios~\cite{CHM,CHM2,CHM3}, the glueball sector  could be involved in causing the spontaneous breaking of the global symmetry responsible for the pseudo-Goldstone Higgs. In this section we explore this possibility.
 
 Let us imagine a sector with a large global symmetry $G$ and a gauge symmetry $G_g$ with no fundamental fermions. Assume then that the dynamics of $G_g$ become strong at some scale $\Lambda$, which then triggers the spontaneous breaking of $G$ down to a smaller subgroup $H$. An example would be a $SO(5)\times SU(3)$ sector, and the $SU(3)$ gauge strong dynamics leading to the breaking of $SO(5)$ to $SO(4)$, see Fig.~\ref{fig:sketch}. One could then proceed as usual in the minimal Composite Higgs scenarios, by partly gauging the $SO(4)$ interactions. To achieve EWSB, one would likely need to use a mechanism as explored in Ref. ~\cite{skyrmions},  where the Higgs potential is generated via a sequential breaking which one could link to the strong sector producing both the spin-zero and -two resonances as well as break the global symmetry. The spectrum at low energies would then contain the Goldstone bosons (a doublet under $SU(2)\subset SO(4)$) and the glueballs.

 In this case, the glueballs will exhibit a hierarchy of couplings. The SM particles with larger couplings to the glueballs would be the Higgs degrees of freedom through a mechanism such as {\it partial compositeness}~\cite{partial-compositeness,PC2,PC3}. As a result, the resonances of the composite sector would couple preferentially to the Higgs doublet, namely, with  the Higgs particle $h$ and the {\it longitudinal} $W^\pm$ and $Z$ bosons. 

Couplings to gluons and photons would be induced through the anomaly terms, and couplings to light fermions would be suppressed by their mass. In Sec.~\ref{sec:couplings} we describe how to parametrize these  couplings. But before we will discuss an alternative view in the context of extra-dimensional theories, which shows the same hierarchical couplings.
 
\subsection{Holographic Radion and Graviton as proxies for Glueballs}

Glueballs can be treated as fields arising from extra-dimensions using dualities. These dualities are based on the AdS/CFT correspondence~\cite{AdSCFTgen}, which draws a duality between strongly coupled theories in $D$ dimensions and a gravitational dual in $D+1$ dimensions, reaching beyond supersymmetric or exactly conformal theories~\cite{LisaPoratti}.  The holographic duality is not an exact mathematical statement but has to be taken as a  qualitative statement between strongly coupled theories (the target theory) and an {\it analogue computer}~\cite{analogue,analogue1,analogue2,analogue3}, a theory on higher dimensions with improved calculability. 

Most of the information of the dual extra-dimensional theory can be understood by means of the metric, which can be expressed as   
\bea
d s^2 = w(z)^2 (\eta_{\mu\nu} d x^\mu d x ^\nu - d z^2)
\eea
with $z$ the extra-dimension and $w(z)$ the warp factor. The extra-dimension is often compactified, $z\in[z_{UV}, z_{IR}]$, with $z_{UV (IR)}$ the UV(IR)-brane. 

Physics at different positions inside the extra-dimension (the bulk) correspond to snapshots of the 4D theory at different energy scales. In other words, the bulk of the extra-dimension encodes the RG evolution of the 4D Lagrangian, with the UV-brane and IR-brane representing the UV and IR boundary conditions on the RGE. Propagation of extra-dimensional fields from the UV to the IR branes correspond to integrating out degrees of freedom: at a position $z_*$ the {\it local} cutoff is related to the UV cutoff as $\Lambda (z_*) = \omega(z_*) \Lambda_{UV}$, a change in cutoff which can be understood a la Wilsonian integrating out heavy degrees of freedom~\cite{lisa-matt}.

The running stops at the IR brane, with the presence of the IR brane signalling that a sector of the 4D theory is undergoing confinement. 4D composite states are therefore dual to the Kaluza-Klein modes due to compactification. Localization of a field in extra-dimensions has a dual meaning as well. Fields localized near or on the UV brane do not strongly participate on the strong dynamics encoded near the IR brane, and are then called {\it elementary}. Localization towards a brane is then the equivalent to the degree of compositeness of the field.   On the other hand, gauge fields in the extra-dimension which are not localized anywhere specifically in the extra-dimension, hence with flat profiles, represent global symmetries of the composite sector, weakly gauged by the UV dynamics~\cite{csaba-fat}. Flat fields are, hence, a mixture of composite and elementary field, much the same as the $\rho-\gamma$ mixing in QCD~\cite{ami-qcd,alex-qcd,analogue}.

In the context of the extra-dimensional theory the roles of the $0^{++}$ and $2^{++}$ states are clear~\cite{real-radion,Witek-Jiji,amif2,GMDM,GMDM2}
\bea
\phi \to \textrm{ radion, and  } G_{\mu\nu} \to \textrm{ KK-graviton} \ ,
\eea
and their behaviour match in both theories in the sense that the structure of couplings of the glueballs and their holographic duals are dictated by symmetries. This dual can be used as a framework to parametrize the properties of the glueballs. Note though that there are differences between the two pictures. The 4D theory does not contain gravity, hence the holographic graviton and radion masses are  unrelated to the scale of quantum gravity, which is assumed to be much higher than the Physics we focus on. Also, terms involving the radion and dilaton at quadratic order may differ~\cite{quadratic-dilaton} when dilatation symmetry is not extended to gravity~\cite{not-grav-dila}. An extensive literature on radion/KK-graviton properties can be found elsewhere, e.g. Ref.~\cite{RSbulk,RSbulk2,RSbulk3,RSbulk4,RSbulk5,RSbulk6,RSbulk7,RSbulk8,RSbulk9,RSbulk10,RSbulk11}.
 
 \subsection{Glueball parametrization}~\label{sec:couplings}
 
 In the partial compositeness picture, larger couplings between the resonances and SM particles indicate a more direct  communication between the new strong dynamics and the SM sector. The hierarchy of couplings we expect is as follows~\cite{diHiggs,GMDM}: {\it 1.) Higgs degress of freedom : } $a_{H}, b_{H} \simeq {\cal O}(1)$, {\it 2.)  Gluon, photon couplings: } $a_{\gamma, g}, b_{\gamma, g} \simeq \alpha_{s,EM}$, and {\it 3.)
Light SM fermions : } $a_f, b_f \propto (m_f/v)^\gamma$, where $\gamma$ is some number, larger than one. 
In the dual extra-dimensional picture, the same hierarchy of couplings has a geometrical meaning~\cite{analogue, diHiggs}. The radion/KK-graviton are bulk states localized near the IR brane, hence with larger overlap with states there, the gluon/photon are delocalized states, and light fermions are bound near the UV brane. Hence, the couplings would go  as above,  see Ref.~\cite{GMDM}.

\section{Glueballs at the LHC} \label{sec3}

In this section we discuss the signatures of glueballs and their interpretation in terms of the reported excess in the diboson channels~\cite{CMSVV,ATLASVV,ATLASVl,CMSVl,ATLASV2l}. This excess is not statistically significant at the moment, hence concrete examples of specific theoretical frameworks are useful to interpret the signatures by performing combinations of different channels. As there is no sensitivity in the fully leptonic channels, and at the same time there is a sizeable overlap among the boosted hadronic $W$, $Z$ and Higgs channels, it turns out that an excess on a channel, e.g. $WZ$, could be in fact due to an excess in the neutral channels. 

In the following, we will assume that the total cross section into dibosons ($WW$ or $ZZ$) is of the order of 1-10 fb, a number to be taken as a ballpark figure. Increasing sensitivity on this excess using the 8 TeV 8 TeV LHCLHC data would require combination of ATLAS and CMS analyses, as well as tailoring to more specific scenarios such as the one presented here.  

Firstly, let us comment on bounds from precision tests of the electroweak sector, which can be parametrized with the help of the S and T parameters~\cite{STUparam}. These bounds on composite $G$ and $\phi$ can be obtained from Refs.~\cite{STR, STG} with suitable modifications. In both cases, the contribution to electroweak observables is induced at loop-level.  Assuming there is no Higgs-$\phi$ mixing term, the contribution to the $S$-parameter due to $\phi$ is~\cite{STR} scales as $S_\phi\propto  \frac{\alpha_H^2 v^2}{\Lambda^2 }$, whereas the spin-two state is $S_G  \approx s_W^2 c_W^2 \frac{b_H^2 m_G^2}{\Lambda^2}$. Both contributions also receive a logarithmic enhancement. As the coupling of $G_{\mu\nu}$ to longitudinal $W$ and $Z$ is through their mass,   $\alpha T$ = 0 at one-loop. 
 
\subsection{Branching ratios and total width}

The partial widths of the $2^+$ state to Higgs degrees of freedom take the following form: $\Gamma( G \to W^+ W^- ) /2\approx \Gamma( G \to Z \, Z )  \approx \Gamma( G \to h \, h ) \approx \frac{b_H^2}{480 \pi} \frac{m_G^3}{\Lambda^2}$,which are correct up to factors of ${\cal O}(m_{Z,W,h}/m_G)^2$. More accurate predictions of the branching ratios can be found at Ref.~\cite{GMDM}. Similarly, the partial width of the $0^+$ state to electroweak bosons are as follows, $\Gamma( \phi \to W^+ W^- ) / 2\approx \Gamma( \phi \to Z \, Z )  \approx   \Gamma( \phi \to h \, h ) \approx \frac{a_H^2}{16 \pi} \frac{m_\phi^3}{\Lambda^2}$.

The  partial width to gluons would be given by $\Gamma(\phi \to g g) = \frac{a_g^2}{2 \pi \Lambda^2} \, m_\phi^3$ and  $\Gamma (G \to g g) = \frac{b_g^2}{120 \pi \Lambda^2} \, m_G^3$.
Therefore, the resonances remain narrow as long as $\Lambda \gtrsim m_{\phi, G}$. Assuming $a_H$ and $b_H$ are the largest couplings, the total width would be given by a simple expression, $\Gamma_\phi \simeq  \frac{a_H^2}{ 4 \pi} \frac{m_\phi^3}{\Lambda^2} $ and $
\Gamma_G \simeq  \frac{b_H^2}{240 \pi} \frac{m_G^3}{\Lambda^2} $.

\subsection{Production cross section}

The production of the glueballs is dominated by gluon fusion (ggF) and possibly vector boson fusion (VBF), as couplings to light fermions are very suppressed. Whether ggF or VBF is the dominant production mechanism depends on the suppression of couplings to gluons respect to $W$ and $Z$ couplings, i.e. the value of the coefficients ($a_{g}$, $b_{g}$) and ($a_{H}$,$b_{H}$) respectively. The production cross section through gluon fusion and VBF is given by the Table below, in the range $m_{\phi,G} = 2 - 1.8$ TeV.
Vector boson fusion is kinematically suppressed, although enhanced by couplings to Higgs degrees of freedom, $a_H$, $b_H$. Note that the VBF 13 TeV LHC is more steep for the spin-two resonance.

\begin{center}
\begin{tabular}{|c||c|c|}
\hline & ggF 8 TeV LHC  & VBF 8 TeV LHC  \\\hline 
$\phi$ & $a_g^2 \, \left(\frac{3 \textrm{ TeV}}{\Lambda}\right)^2$ (370 - 740) fb & $a_g^2 \, \left(\frac{3 \textrm{ TeV}}{\Lambda}\right)^2$ (0.002-0.004) fb  \\\hline 
$G$ & $b_g^2 \, \left(\frac{3 \textrm{ TeV}}{\Lambda}\right)^2$ (90 - 190) fb & $b_g^2 \, \left(\frac{3 \textrm{ TeV}}{\Lambda}\right)^2$ (0.6 - 1.3) fb  \\\hline
&ggF 13 TeV LHC & VBF 13 TeV LHC 
 \\\hline 
 $\phi$ & $a_g^2 \, \left(\frac{3 \textrm{ TeV}}{\Lambda}\right)^2$ (5 - 8) pb & $a_H^2 \, \left(\frac{3 \textrm{ TeV}}{\Lambda}\right)^2$(0.03-0.05) pb
  \\\hline
  $G$ & $b_g^2 \, \left(\frac{3 \textrm{ TeV}}{\Lambda}\right)^2$ (1 - 2) pb & $b_H^2 \, \left(\frac{3 \textrm{ TeV}}{\Lambda}\right)^2$ (0.02 - 0.03) pb
  \\\hline 
\end{tabular}
\end{center}
 
\begin{figure}[t!]
\begin{center}
\includegraphics[height=.25\textheight]{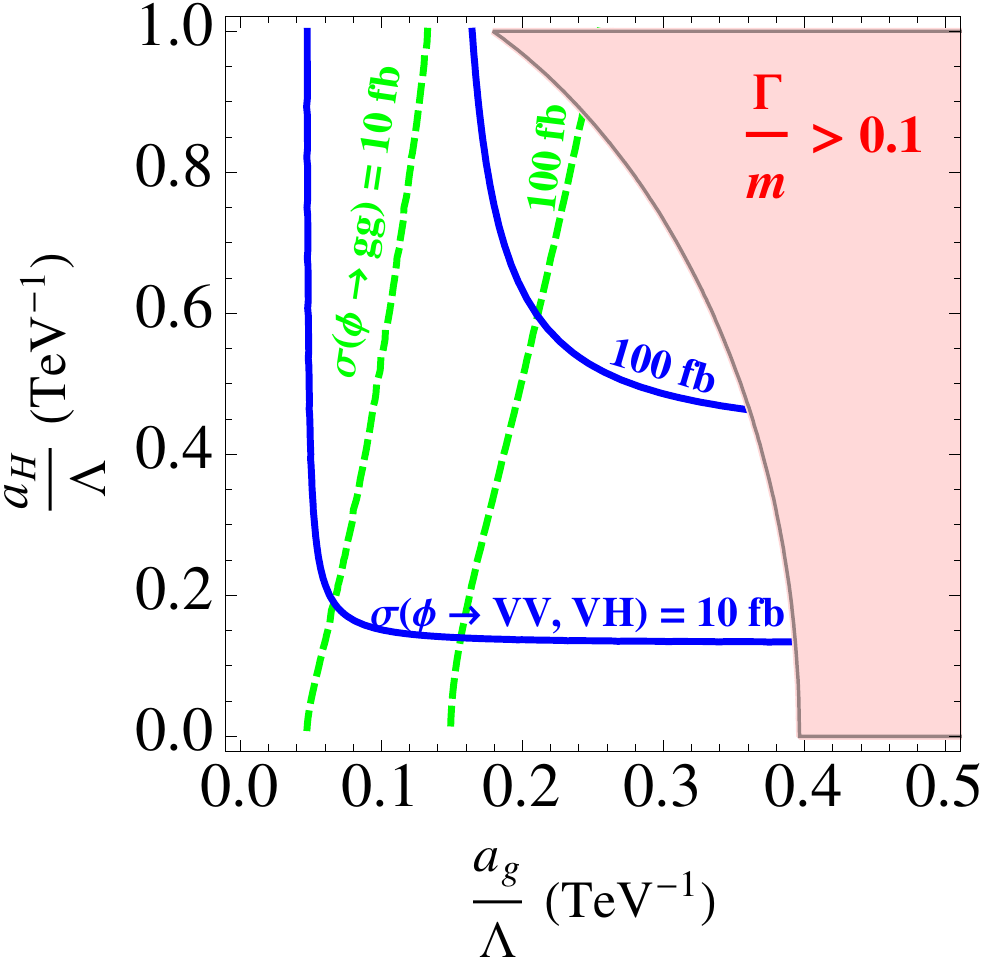}
\hspace{0.5cm}
\includegraphics[height=.25\textheight]{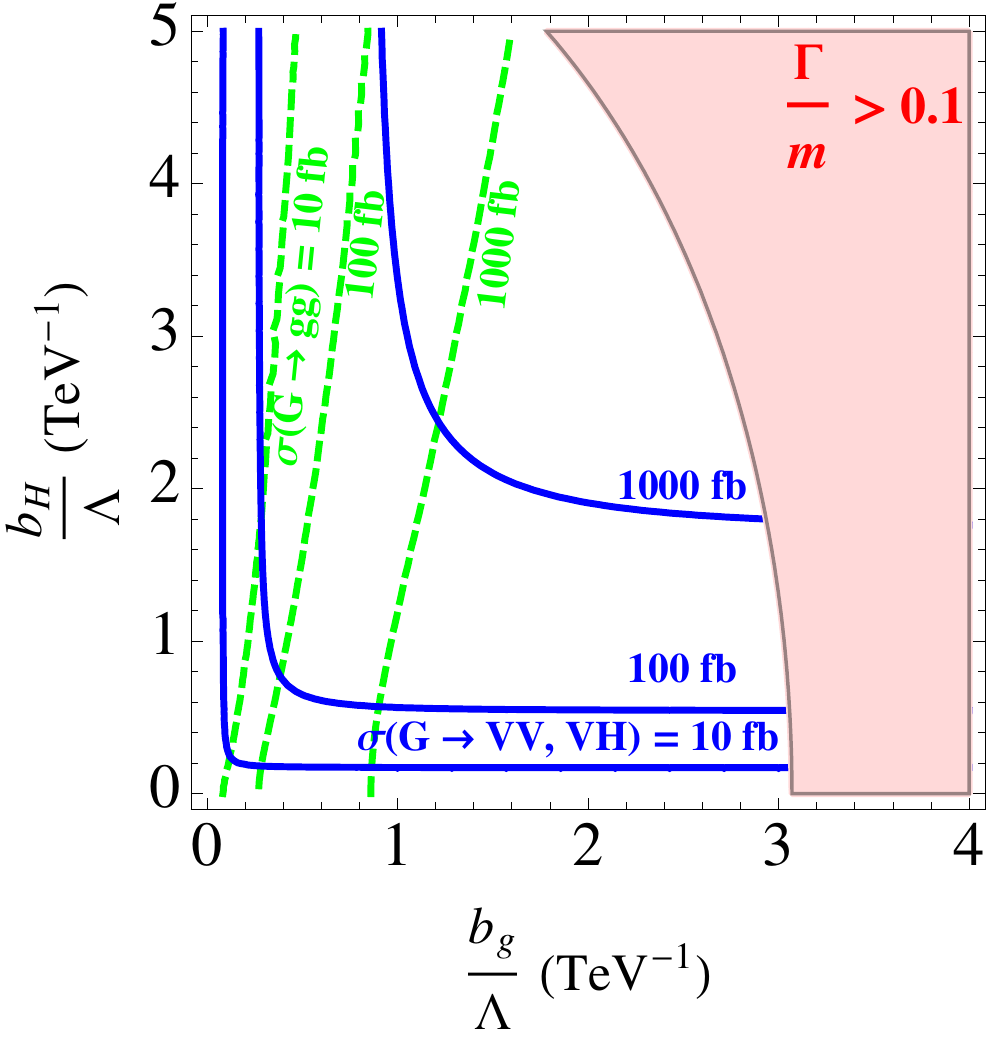}
\label{fig2}
\caption{Total cross sections at 8 TeV LHC for the spin-zero (left) and -two (right) hypothesis in the $HV$ and $VV$ channels and di-gluon final state. Note cross sections do not include efficiencies to cuts.}
\end{center}
\end{figure}

 In Fig.~\ref{fig2} we show the total cross sections at 8 TeV LHC for the spin-zero and -two hypothesis in the $HV$ and $VV$ channels and di-gluon final state. As one can see, the spin-two resonance remains narrow in a larger region of the parameter space, a behaviour which is related to the kinematic features discussed in the next section. These numbers have been obtained by implementing the glueballs in the context of {\tt feynrules}~\cite{feynrules}, exported to {\tt Madgraph5}~\cite{MG5} using the {\tt UFO}~\cite{UFO} format. 
 
The coupling of the glueball $\phi$ to gluons, photons and vector bosons has the same Lorentz structure as the Higgs (see Eqs.~\ref{r1} and ~\ref{r2}), whereas the spin-two resonance has a more interesting Lorentz structure. In particular, the Feynman rule (in the unitary gauge) involving this resonance with vector bosons is given by~\cite{GMDM}
\bea
&&\left[G_{\mu\nu}, V_\alpha(k_1) , V_\beta(k_2) \right]:-i \frac{1}{\Lambda} ( b_H m^2_V C_{\mu\nu,\alpha\beta}+b_V W_{\mu\nu,\alpha\beta}) \, \eea
where 
\bea
W_{\mu\nu,\alpha\beta} &\equiv&\eta_{\alpha\beta}k_{1\mu}k_{2\nu}+\eta_{\mu\alpha}(k_1\cdot k_2\,\eta_{\nu\beta}-k_{1\beta}k_{2\nu} )-\eta_{\mu\beta} k_{1\nu}k_{2\alpha}+ \nonumber \\ 
& & \frac{1}{2}\eta_{\mu\nu}(k_{1\beta}k_{2\alpha}-k_1\cdot k_2\, \eta_{\alpha\beta}  )+(\mu \leftrightarrow \nu) , \nonumber \\
C_{\mu\nu,\alpha\beta} &\equiv&\eta_{\mu\alpha}\eta_{\nu\beta}+\eta_{\nu\alpha}\eta_{\mu\beta}-\eta_{\mu\nu}\eta_{\alpha\beta} \ ,
\eea
and $V$ here denotes either a massive vector boson ($b_H\neq 0$) or a massless gluon or photon ($b_H=0$ and $b_V=b_{g,\gamma}$).

\begin{figure}[h!]
\begin{center}
\includegraphics[width=.5\textwidth]{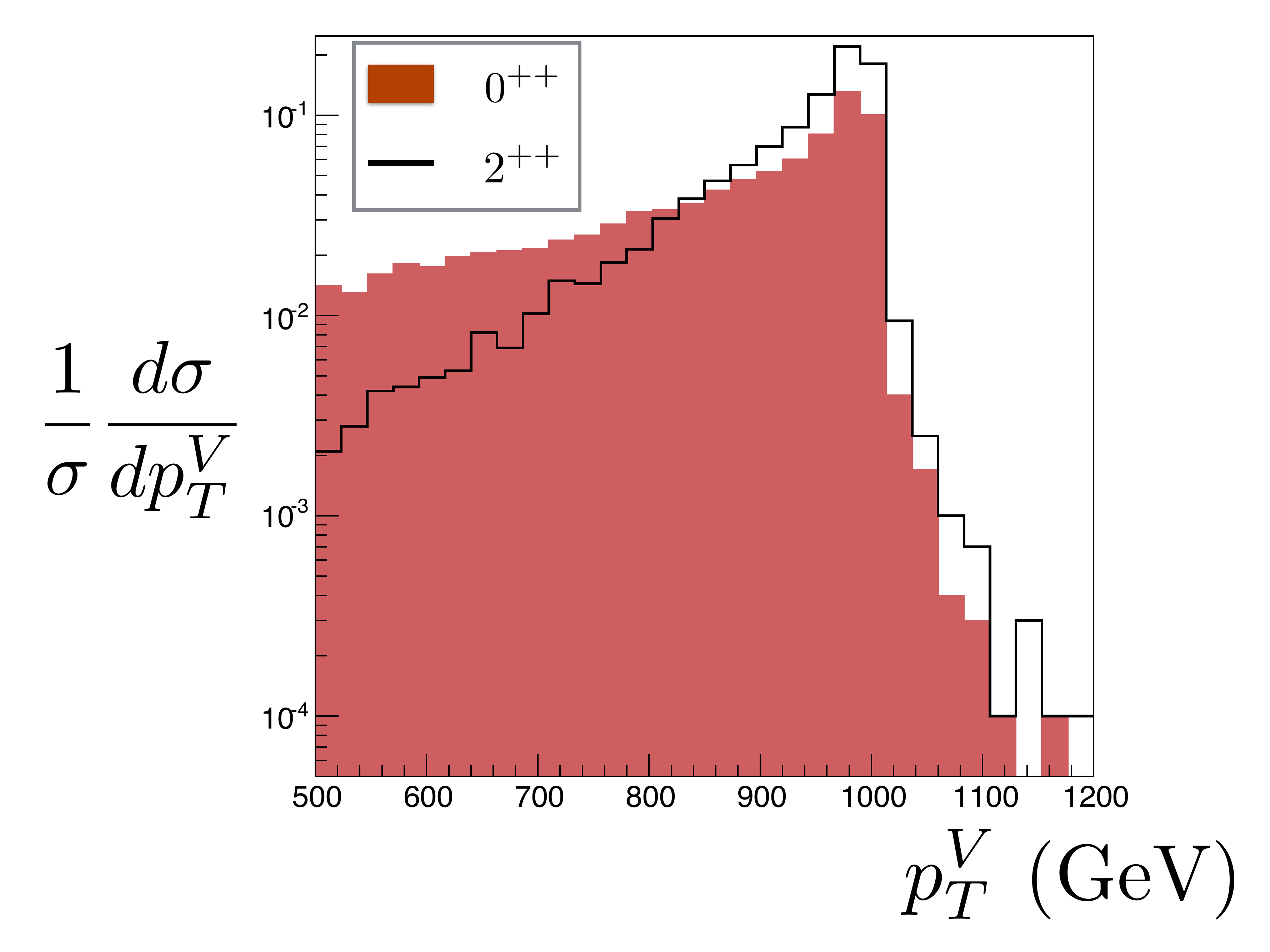}
\label{ptV}
\caption{Distribution of the transverse momentum of the leading vector boson in the decay of $0^{++}$ and $2^{++}$ resonances at 8 TeV LHC.}
\end{center}
\end{figure}

These differences between the structure of couplings of $\phi$ and $G_{\mu\nu}$ is manifested not only in the total cross section and the ratio of ggF versus VBF discussed in the last section. Kinematic distributions, hence acceptance to cuts and reconstruction will be modified. As an illustration, we show in Fig.~\ref{ptV} the $p_T$ distribution of the leading V-jet for both spin hypothesis. Note that this simulation has been done at parton-level,  and showering/hadronization and detector effects would distort the distribution.

The higher spin resonance exhibits are more boosted spectrum due to the different Lorentz structure of the coupling, see Ref.~\cite{withJohn,withJohn2,withJohn3,withJohn4,withJohn5} for a discussion in the context of the Higgs-candidate.  Therefore, criteria such as mass drop or mass reconstruction will depend on the spin of the resonance. The spin of the resonance determines not only the $p_T$ spectrum but also angular distributions. We leave for a future study~\cite{inprep} a more realistic analysis of this scenario within a feasibility study at Les Houches.
 
  \section{Discussion of the results}  \label{concl}
  
 The Run1 data has been able to probe diboson production in the very boosted regime, providing a tantalizing hint of new physics at around 2 TeV. How fast and reliably this hint can be confirmed or discarded depends on contrasting the signatures with specific models. Combination of different diboson channels, as well as with dijet, di-top and Higgs-vector boson relies on benchmarks against which the analyses can be tested. Besides trivial issues of branching ratios to specific final states, the analysis depends on whether a resonance is gluon, quark or vector boson initiated, and its quantum numbers, as they determine how the decay products are distributed.  
   
In this paper we have such benchmark of  interpretation in the context of glueballs of a new strong sector. We have discussed how the lightest states, massive spin-zero and spin-two resonances, could inherit their couplings from the breaking of diffeomorphism invariance due to the strong dynamics. They would then behave in a similar way as expected from a radion and a massive Kaluza-Klein resonance of extra-dimensions. 

If the new sector dynamics is linked to EWSB, the resonances will couple preferentially to Higgs degrees of freedom, i.e. the Higgs and longitudinal polarizations of the $W$ and $Z$ bosons, whereas the best way to produce them is through gluon fusion with a possible vector-boson fusion component. 

Regarding the collider analysis, the results presented here are performed at parton level. A more detailed analysis, including prospects to measure properties with 13 TeV LHC will be done in a future publication~\cite{inprep}.
  
 \section*{Note added}
 After this paper was written, new results from the Run2 LHC looking for the diboson resonance show no significant excess in the region around 2 TeV. More data in the summer 2016 would be required to determine whether the Run1 excess is ruled out by a larger Run2 dataset. 
  
\section*{Acknowledgements}
I would like to thank Gustaaf Broojimans and Chris Pollard for discussions on the diboson analyses, as well as the excellent atmosphere in Les Houches. Many thanks to Axel Maas and Agostino Patella for discussions on the spectrum of glueballs.
This work  is supported by the Science Technology and Facilities Council (STFC) under grant number ST/L000504/1.

\end{document}